\documentclass{osa-article}

\newcommand{\ts}{\textsuperscript}

\journal{oe}

\articletype{Research Article}

\begin{document}


\title{Demonstration of second harmonic generation in gallium phosphide nano-waveguides}

\author{Aravind P. Anthur,\authormark{1,3} Haizhong Zhang,\authormark{1,3} Yuriy Akimov,\authormark{2,3} Jun Rong Ong,\authormark{2,3} Dmitry Kalashnikov,\authormark{1} Arseniy I. Kuznetsov,\authormark{1} and Leonid Krivitsky\authormark{1,*}}

\address{\authormark{1}Institute of Materials Research and Engineering (IMRE), A*STAR (Agency for Science, Technology and Research) Research Entities, 2 Fusionopolis Way, $\#$08-03 Innovis, Singapore 138634.\\
\authormark{2}Institute of High Performance Computing (IHPC), A*STAR (Agency for Science, Technology and Research) Research Entities, 1 Fusionopolis Way, $\#$16-16 Connexis, Singapore 138632.\\
\authormark{3}Equal contribution}

\email{\authormark{*}leonid\_krivitskiy@imre.a-star.edu.sg} 



\begin{abstract*}
We designed, fabricated and tested gallium phosphide (GaP) nano-waveguides for second harmonic generation (SHG). We demonstrate SHG in the visible range around 655 nm using low power continuous-wave pump in the optical communication O-band. Our structures utilize modal phase matching, such that lower order eigenmodes of the pump are phase matched to higher order eigenmodes of the second harmonic. We observe phase matched SHG for different combinations of interacting modes by varying the widths of the waveguides and tuning the wavelength of the pump. The presented results contribute to the development of integrated photonic platforms with efficient nonlinear wave-mixing processes for classical and quantum applications. 
\end{abstract*}

\section{Introduction}
Integrated photonics promises miniaturization of optical devices, leading to the development of scalable on-chip solutions \cite{Fathpour_2018, Minzioni_2019}. One of the key challenges of integrated quantum photonics is that the most efficient detection, storage and manipulation of photons are done in the visible spectral range, whereas the most efficient transmission of photons occurs at the telecom wavelengths. Nonlinear optical processes can be exploited to bridge this gap \cite{Greve_2012, Zaske_2012}. The nonlinear material for the frequency conversion process should satisfy several requirements: transparency at both the telecom and the visible wavelengths, high nonlinearity and high refractive index to provide strong light confinement for compact devices on integrated photonic platforms. 

A number of integrated platforms have been developed so far based on different nonlinear materials like lithium niobate (LiNbO$_3$), gallium arsenide (GaAs), aluminium nitride (AlN) and gallium nitride (GaN) \cite{Chowdhury_2000, Rao_2016, Wang_2017, Wang_2017OE, Ge_2018, Chen_2018, Pernice_2012, Bruch_2018, Rao_2004, Rutkowska_2011, Chowdhury_2003, Roland_2016, Rigler_2018 }. At present, the most promising bulk material for commercial applications is LiNbO$_3$ because of its broad transparency window, high $\chi^{(2)}$, and the possibility of periodic domain inversion \cite{Wang_2017Thesis}. However, nanofabrication of LiNbO$_3$ is challenging and hence recent demonstrations of LiNbO$_3$ nanostructures depend on materials like silicon nitride (SiN) to provide waveguiding \cite{Chang_2016}. Also, its relatively low refractive index of of $\sim$2.2 makes dense integration of nanophotonic devices difficult. Semiconductor materials like AlGaAs and GaAs have even larger nonlinear parameter $\chi^{(2)}$ and higher refractive index \cite{Rao_2004}. However, the transparency cut-off wavelength for AlGaAs is around 900 nm \cite{Wilson_2019}. Hence AlGaAs cannot be used in the visible wavelength range which is relevant to applications in imaging, quantum optics and sensing. AlN is another material which is under consideration \cite{Bruch_2018}. However, AlN has relatively low refractive index and $\chi^{(2)}$. A comparison of the material parameters is given in Table \ref{tbl:MaterialParam}. 

For these reasons, we have selected gallium phosphide (GaP) as a promising candidate for frequency-conversion. It has a high refractive index ($\sim$3.1 in the O-band), high second-order nonlinear parameter (d$_{36}$ of $\sim$50 pm/V in the O-band), good thermal conductivity for temperature tuning and broad transparency range from 550 nm to 11 um \cite{Ziel_1975, Rivoire_2009, Shambat_2010, Rivoire_2011, Helmy_2011, Sanatinia_2012, Vaclavik_2013, Sanatinia_2014, Guha_2015, Sanatinia_2015, Lake_2016, Ye_2017, Cambiasso_2017, martin_2018nonlinear, Logan_2018, Schneider_2018, Dave_2018, Poulvellarie_2019, Wilson_2019}. As a first step in developing an integrated GaP platform for frequency conversion, we target to demonstrate second harmonic generation (SHG) in nano-waveguides. SHG remains the most studied and widely applied second-order nonlinear process among others \cite{Boyd_NLO}, for a broad variety of applications, including lasers, pulse characterization, quantum optics, spectroscopy and imaging to name a few \cite{Armstrong_1967,Heinz_1982, Freund_1986, Rottwitt_NLO:PandA}. Hence, more efficient platforms for SHG are always desirable and motivates a strong interest in the community for studying nonlinear frequency mixing processes. 

GaP is gaining interest in the research community \cite{martin_2018nonlinear,Wilson_2019}, but thus far it is relatively less well studied compared to other nonlinear materials, with few articles describing $\chi^{(2)}$ processes in waveguides. In this work, we first review the theory of SHG in GaP waveguides. The crystal axis orientation is of crucial importance in design and modeling of the waveguides, as it is required to calculate the overlap integrals for the mode coupling in SHG. We thus measured and confirmed the GaP crystal axis orientation of the thin film on our samples. With the developed formalism, we further elaborate on SHG in nano-waveguides by modal phase matching. We then describe the nano-waveguide fabrication process using samples of thin film GaP on SiO$_2$, on top of sapphire substrate. Finally, we report our experimental results on SHG in GaP nano-waveguides.   

\begin{table}[tb]
	\captionsetup{font={small,bf}}
	\centering
	\caption{Parameters of nonlinear materials at 1310 nm}
	\label{tbl:MaterialParam}
	\footnotesize
	\begin{tabular}[c]{| c | c | c | c |}
		\hline
		\textbf{Material}	& \textbf{Linear refractive index \cite{Bond_1965,RefractiveIndexInfo}}	& \textbf{d$_{\textit{eff}}$ (pm/V) \cite{Shoji_1997,Bruch_2018,Rigler_2018}} & \textbf{Cut-off $\lambda$ (nm) \cite{Wilson_2019}} \\ \hline
		\hline
		AlN & 2.124 & 6 & 200 \\
		\hline
		GaN & 2.323 & 6 & 400 \\
		\hline
		LiNbO$_3$ & 2.220 & 20 & 310 \\
		\hline 
		GaP & 3.074  & 50 & 550 \\
		\hline
		GaAs & 3.404 & 110 & 900 \\
		\hline
	\end{tabular}
\end{table}

\section{SHG in GaP waveguides}

The fields excited in the waveguide at the angular frequency $\omega$ can be given as a superposition of eigenmodes \cite{Yariv_1973},
\begin{equation}
\textbf{E}(\textbf{r},\omega) = \sum_m A_{m}(z,\omega) \tilde{\textbf{E}}_{m}(x,y,\omega)e^{i\beta_m(\omega) z},
\end{equation}
where $\textbf{E}(\textbf{r},\omega)$  is the electric field, $m$ is the eigenmode index, $A_m (z,\omega)$ is the complex amplitude of the $m$\ts{th} eigenmode along the propagation direction $(z)$, $\tilde{\textbf{E}}_m (x,y,\omega)$ is the spatial distribution of the electric eigen field in the waveguide cross-section plane, and $\beta_m (\omega)$ is the propagation constant of the $m$\ts{th} eigenmode. The eigen fields $\tilde{\textbf{E}}_m (x,y,\omega)$  are normalized by the condition,
\begin{equation}
-\frac{i}{2\mu_0\omega}\iint \textbf{e}_z\cdot\left(\tilde{\textbf{E}}^*_m(x,y,\omega) \times \left[\nabla \times \tilde{\textbf{E}}_m(x,y,\omega) \right]\right)dxdy = 1 ~ [W],     
\end{equation}
that corresponds to 1 W of total power transferred by the $m$\ts{th} eigenmode in the $z$-direction, where $\mu_0$ is the vacuum permeability. Then, interactions between eigenmodes at frequency $\omega$ and $2\omega$ are governed by the coupled equations,

\begin{equation}
\frac{dA_m(z,\omega)}{dz} = \sum_{n,l} \kappa_{mnl}(2\omega,-\omega)A_n^*(z,\omega)A_l(z,2\omega)e^{-i \Delta \beta_{mnl}(\omega)z},
\label{eq:pump}
\end{equation}
\begin{equation}
\frac{dA_l(z,2\omega)}{dz} = - \sum_{m,n} \kappa^*_{mnl}(\omega,\omega)A_m(z,\omega)A_n(z,\omega)e^{i \Delta \beta_{mnl}(\omega)z},
\label{eq:SHG}
\end{equation}
with the three eigenmode phase mismatch, $\Delta \beta_{mnl}(\omega)$, given by,
\begin{equation}
\Delta \beta_{mnl}(\omega) = \beta_m(\omega) + \beta_n(\omega) - \beta_l(2\omega),
\label{eq:delta_beta}
\end{equation}
and the overlap integrals,
\begin{equation}
\begin{split}
\kappa_{mnl}(2\omega,-\omega) = i\frac{\omega\epsilon_0}{4}\sum_{i,j,k}\iint \chi^{(2)}_{ijk}(x,y,\omega,2\omega)\left[\tilde{\textbf{E}}^*_m(x,y,\omega)\cdot\textbf{e}_i\right]\left[\tilde{\textbf{E}}^*_n(x,y,\omega)\cdot\textbf{e}_j\right]\\ \times \left[\tilde{\textbf{E}}_l(x,y,2\omega)\cdot\textbf{e}_k\right]dxdy ~ [W^{-1}],
\end{split}
\label{eq:kappa1}
\end{equation}
\begin{equation}
\begin{split}
\kappa_{mnl}(\omega,\omega) = i\frac{(2\omega)\epsilon_0}{4}\sum_{i,j,k}\iint \chi^{(2)*}_{ijk}(x,y,\omega,\omega)\left[\tilde{\textbf{E}}^*_m(x,y,\omega)\cdot\textbf{e}_i\right]\left[\tilde{\textbf{E}}^*_n(x,y,\omega)\cdot\textbf{e}_j\right]\\ \times \left[\tilde{\textbf{E}}_l(x,y,2\omega)\cdot\textbf{e}_k\right]dxdy ~ [W^{-1}].
\end{split}
\label{eq:kappa2}
\end{equation}
The second-order susceptibility tensors, $\chi^{(2)}_{ijk}$, that define the overlap integrals strongly depends on the waveguide composition. We consider GaP waveguides embedded in silicon dioxide cladding. The tensor components $\chi^{(2)}_{ijk}(x,y)=0$ when the coordinates $(x,y)$ are outside the GaP waveguide. Under the Kleinman symmetry conditions, $\chi_{ijk}^{(2)} (\omega,2\omega) = \chi^{(2)}_{ijk}(\omega,\omega) \equiv \chi^{(2)}_{ijk}$. Also, the zinc blende crystal structure of GaP further reduces the total number of tensor elements that need to be considered: $\chi_{xyz}^{(2)} = \chi_{xzy}^{(2)} = \chi_{yzx}^{(2)} = \chi_{yxz}^{(2)} = \chi_{zxy}^{(2)} = \chi_{zyx}^{(2)} = 2 d_{36}$ \cite{Boyd_NLO}. In contracted matrix notation, the second-order nonlinear susceptibility tensor for GaP is
\begin{equation}
d=
\begin{pmatrix}
0 & 0 & 0 & d_{36} & 0 & 0 \\
0 & 0 & 0 & 0 & d_{36} & 0 \\
0 & 0 & 0 & 0 & 0 & d_{36} \\
\end{pmatrix}
\label{eq:dmatrix}.
\end{equation}
We assume $d_{36}$ = 50 pm/V in our analysis \cite{Levine_1972,Shoji_1997}. 

As the crystal axis is not necessarily the same as the waveguide axis, it is important to do the rotational transformation between the two coordinates when calculating the overlap integrals. In our case the GaP crystal axis orientation, as stated by the vendor, has to be tilted with respect to the sample surface. Confirmation of the tilt angle is needed to calculate the overlap integrals and is critical to guide the design of the nano-waveguides \cite{Kuo_2018, Dave_2018}. We confirmed the rotation angle in a separate experiment, the details of which are given in the next section.

\subsection{Measurement of crystal axis orientation}

The surface normal vector of the thin film gallium phosphide (GaP) sample is tilted at an angle $\theta = 15^\circ$ towards the [\textbf{111}] crystal direction. This tilt angle was chosen by the vendor because it is helpful for the nucleation in hetero-epitaxial growth with large lattice mismatch. More details on the fabrication flow are shown in Section \ref{sec:fab}. We verify the tilt angle by measuring the SHG under normal incident pump. We rotate the GaP thin film around the normal direction and compare the experimentally measured SHG with the simulation data for various tilt angles $\theta$.  

\begin{figure}
	\centering
	\includegraphics[width=0.75\linewidth]{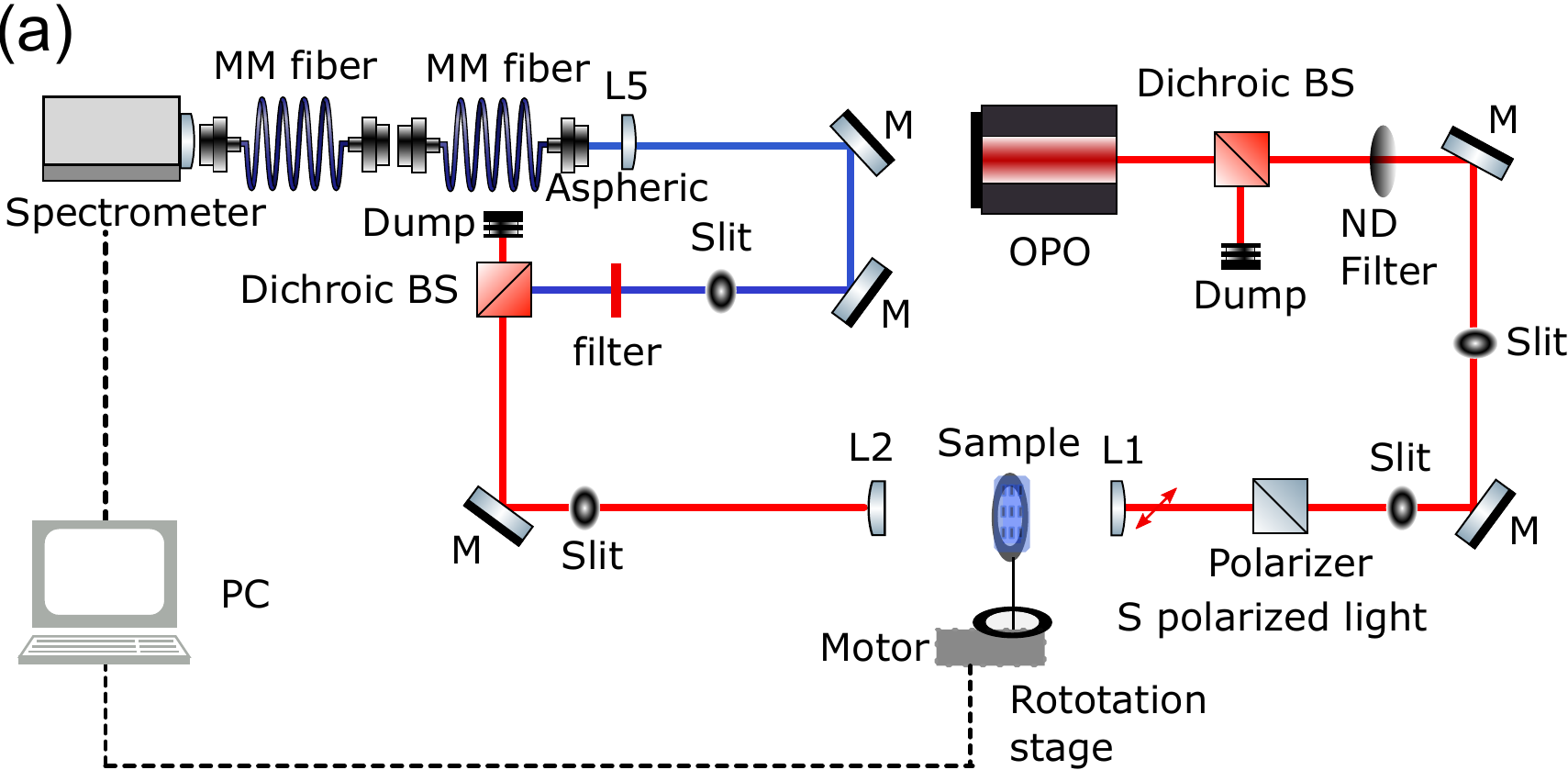}
	\includegraphics[trim={80 80 90 100},clip,width=0.85\linewidth]{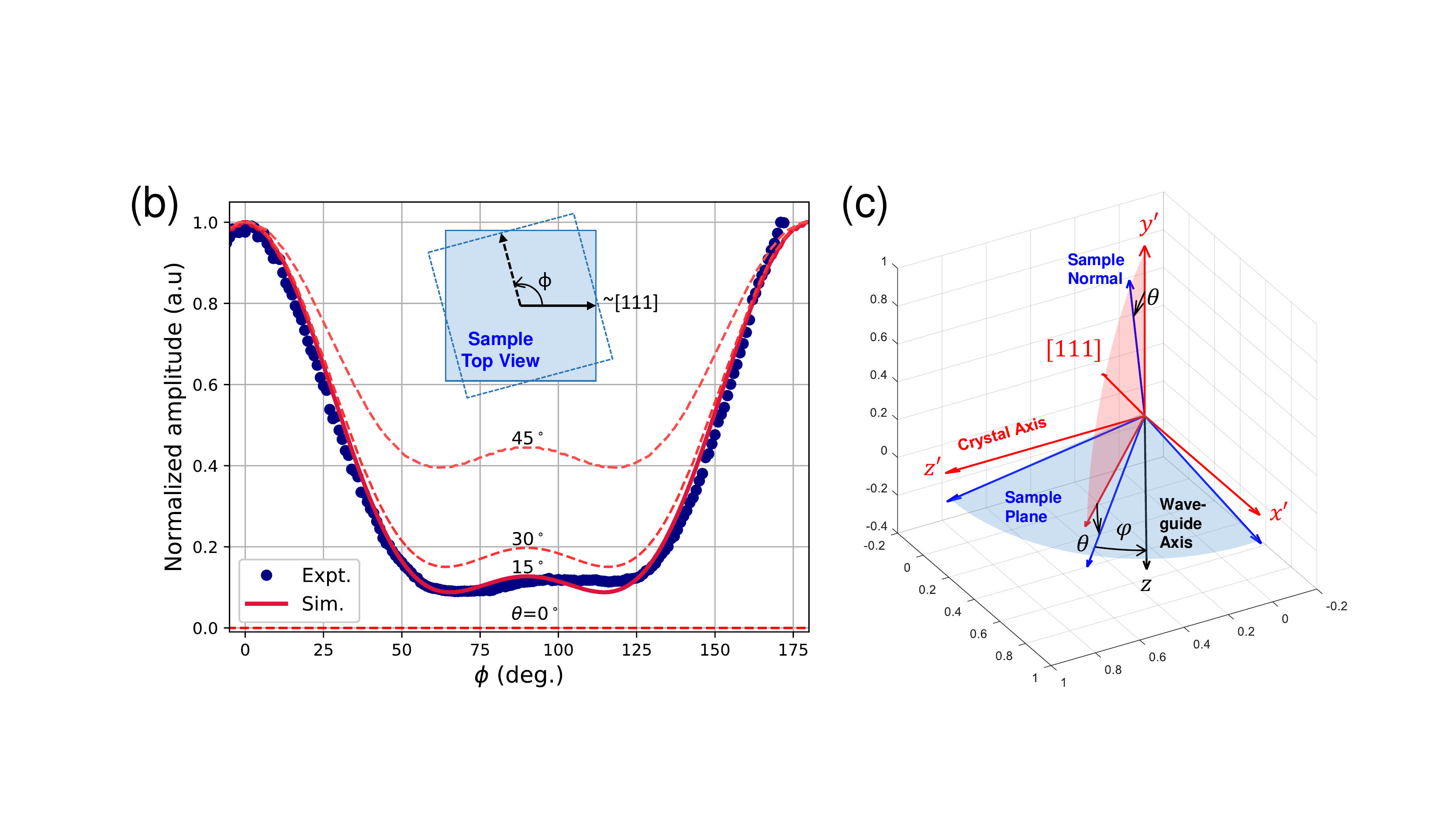}
	\caption{(a) Experimental system used to analyze the crystal axis of GaP sample. MM--multi-mode, M--mirror, BS--beam splitter, L--lens, OPO--optical parametric oscillator. (b) Experimental (blue dots) and simulated normalized SHG amplitude (red lines) as a function of the rotation angle $\phi$ of the sample. $\theta = 15^\circ$ matches best with the experiment data. (c) Visualization of the relationships between the crystal axis, the sample plane and the waveguide axis. }
	\label{fig:Experimental_crystalaxis}
\end{figure}

Figure \ref{fig:Experimental_crystalaxis}(a) shows the schematic of the experimental setup \cite{Optics}. The pump is a pulsed optical parametric oscillator (OPO) at 1310 nm, pumped by a Ti-Sapphire oscillator with the spectral width of 11 nm and repetition rate of 78 MHz. Dichroic beam splitter is used to spectrally filter the pump pulses from the OPO at 1310 nm. The light from the output of the dichroic passes through a polarizer to increase the extinction ratio of the polarized output from the OPO. The light then passes through a plano-convex lens and focusses at the thin film sample of GaP. The objective lens with an NA of 0.45 then collects the output from the thin-film. The light at the output of the thin film GaP sample contains both the pump and the SHG signal. A dichroic beam splitter and filter are used to filter the SHG from the sample at the wavelength of approximately 655 nm. The filtered SHG signal is then collected and analyzed using a spectrometer. 

The sample is rotated over 180 degrees around the normal, with normally incident pump at horizontal (H) polarization. The strength of the SHG signal is analyzed as a function of the rotation angle $\phi$, which is the angle between the pump polarization and the projection of [\textbf{111}] onto the sample plane. This projected vector is perpendicular to the edge of the sample and for $\phi = 0^\circ$ it is parallel to the pump polarization. Figure \ref{fig:Experimental_crystalaxis}(b) shows the experiment results (blue dots). We expect that there should not be any SHG observed if any of the family of $\langle$\textbf{100}$\rangle$ crystal directions coincides with the normal to the wafer. This is because the only non-vanishing elements in the nonlinear susceptibility tensor of GaP mix three different field components. The observation of SHG for normally incident pump is proof that the surface normal is tilted relative to the crystal axis. Comparing with the simulation results (red lines), $\theta = 15^\circ$ has the best match, verifying the crystal axis tilt angle. A clear signature of the tilt angle $\theta$ is the ratio of SHG powers, $\frac{P_{2\omega}(\phi=0^\circ)}{P_{2\omega}(\phi=90^\circ)}$, which decreases as $\theta$ increases. Note that there is a small shift of $\phi$ by approximately $3^\circ$ between the simulation and the experiment, that is compensated in Fig. \ref{fig:Experimental_crystalaxis}(b), due to the rotation of the $0^\circ$ definition in the experimental system. 

To calculate the overlap integrals, we fix the sample frame and allow the waveguides to rotate in the sample plane. Thus, in the following section, we define $\varphi$ as the angle between the waveguide axis and the projection of [\textbf{111}] onto the sample plane. Figure \ref{fig:Experimental_crystalaxis}(c) visualizes the relationships between the various reference frames.

\subsection{Modal phase matching}

For SHG in waveguides, the exact phase matching condition for two eigenmodes is  $n_{2\omega} = n_{\omega}$. This means that the pump mode and second harmonic mode effective indices should be equal. Due to material dispersion, the effective index of same order eigenmodes increases as the wavelengths decreases. On the other hand, effective index decreases with increasing mode order. As such, modal phase matching is achieved by adjusting the waveguide dimensions so that a higher order eigenmode of the second harmonic has the same effective index as a lower order eigenmode of the pump. 

\begin{figure}
	\centering
	\includegraphics[trim={0 150 0 100},clip,width=\linewidth]{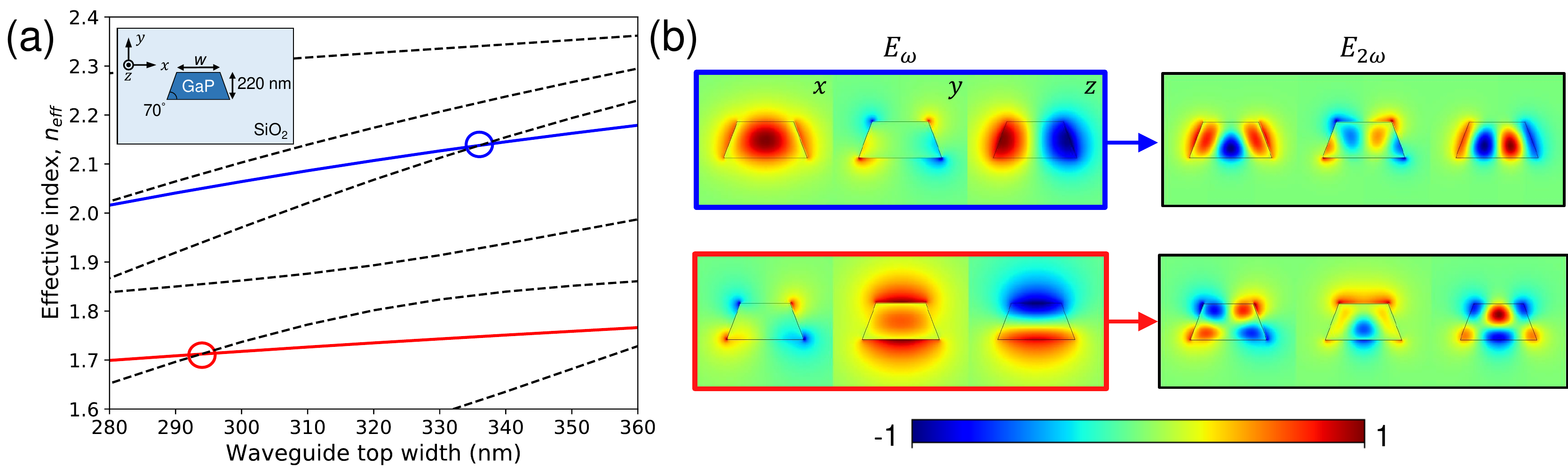}
	\caption{(a) Effective index plot versus top width of GaP waveguide, at 1310 nm. The phase matched points are circled. The inset shows the geometry of the GaP waveguides. (b) The electric field components of the two lowest order pump eigenmodes (red and blue) and the corresponding phase matched higher order eigenmode of the second harmonic.}
	\label{fig:Modes}
\end{figure}

Figure \ref{fig:Modes}(a) shows the calculated mode effective indices for a waveguide height of 220 nm and side wall angle of 70$^{\circ}$, at a pump wavelength of 1310 nm. The dotted lines are the indices of eigenmodes at the second harmonic wavelength and the red and blue colored lines are the indices of the two lowest order pump eigenmodes. Figure \ref{fig:Modes}(b) shows the simulated electric field component profiles for each of these. The 1\ts{st} mode (blue), commonly called a quasi-TE mode, has the largest components in the $x$ and $z$ direction and is excited by injecting horizontal (H) polarized pump light. The 2\ts{nd} mode (red), commonly called a quasi-TM mode, has the largest components in the $y$ and $z$ direction and is excited by injecting vertical (V) polarized pump light. A common feature of these two pump eigenmodes is the two lobed $E_z$ fields. The second-order nonlinear susceptibility tensor of GaP enables unique modal coupling configurations for these dipolar eigenmodes \cite{Poulvellarie_2019}. 

We identify phase matched configurations based on the line intersections, circled in Fig. \ref{fig:Modes}(a). The quasi-TE (blue) and quasi-TM (red) modes are each separately phase matched with higher order second harmonic modes. For SHG, the coupled equations become
\begin{equation}
\frac{dA_\omega(z)}{dz} = \kappa A_\omega^*(z)A_{2\omega}(z)e^{-i \Delta \beta z},
\label{eq:pump-pm}
\end{equation}
\begin{equation}
\frac{dA_{2\omega}(z)}{dz} = - \kappa^* A_\omega(z)A_\omega(z)e^{i \Delta \beta z},
\label{eq:SHG-pm}
\end{equation}
with $\Delta \beta = 2 \beta_\omega - \beta_{2\omega}$. In general, the crystal axis $(x',y',z')$ need not coincide with the waveguide axis $(x,y,z)$. The vector components of the electric fields, originally described in the waveguide frame, have to be recast into the crystal frame prior to performing the overlap integral. After accounting for the GaP crystal symmetry, the overlap integral is now, 
\begin{equation}
\begin{split}
\kappa =~ & i\frac{\omega\epsilon_0}{2} \iint 2 d_{36} \times \\ 
\bigg[ & 
\Big(\sum_{q} R_{x'q} \cdot \tilde{\textbf{E}}_{2\omega}(x,y)\cdot\textbf{e}_q\Big)
\Big(\sum_{q} R_{y'q} \cdot \tilde{\textbf{E}}^*_{\omega}(x,y)\cdot\textbf{e}_q\Big)
\Big(\sum_{q} R_{z'q} \cdot \tilde{\textbf{E}}^*_{\omega}(x,y)\cdot\textbf{e}_q\Big) \\ 
+ & 
\Big(\sum_{q} R_{y'q} \cdot \tilde{\textbf{E}}_{2\omega}(x,y)\cdot\textbf{e}_q\Big)
\Big(\sum_{q} R_{z'q} \cdot \tilde{\textbf{E}}^*_{\omega}(x,y)\cdot\textbf{e}_q\Big)
\Big(\sum_{q} R_{x'q} \cdot \tilde{\textbf{E}}^*_{\omega}(x,y)\cdot\textbf{e}_q\Big) \\ 
+ &
\Big(\sum_{q} R_{z'q} \cdot \tilde{\textbf{E}}_{2\omega}(x,y)\cdot\textbf{e}_q\Big)
\Big(\sum_{q} R_{x'q} \cdot \tilde{\textbf{E}}^*_{\omega}(x,y)\cdot\textbf{e}_q\Big)
\Big(\sum_{q} R_{y'q} \cdot \tilde{\textbf{E}}^*_{\omega}(x,y)\cdot\textbf{e}_q\Big) \bigg] dxdy.
\end{split}
\label{eq:kappa-sim1}
\end{equation}
with the integrand being non-zero only within the waveguide. $\textbf{R}$ is the rotation matrix such that $\tilde{\textbf{E}}' = \textbf{R}\cdot \tilde{\textbf{E}}$ and $q$ are the waveguide axes $(x,y,z)$. The strength of the coupling between the interacting eigenmodes is determined by the mode overlap $\kappa$. Efficient SHG can occur if the conditions of modal phase matching and good mode overlap are simultaneously fulfilled.

\begin{figure}
	\centering
	\includegraphics[trim={0 0 20 20},clip,width=0.85\linewidth]{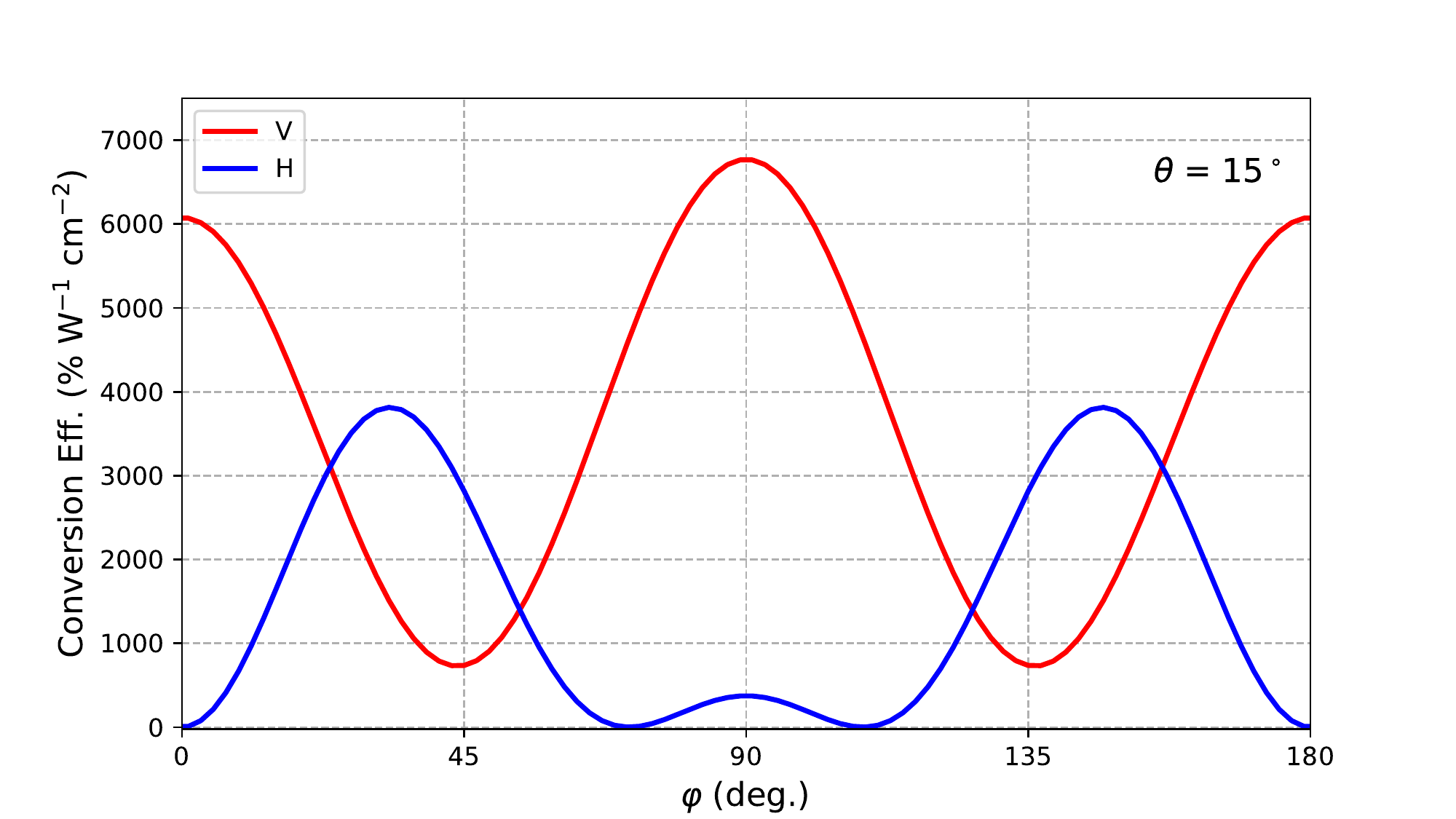}
	\caption{Calculated normalized conversion efficiency ($\%$ W$^{-1}$cm$^{-2}$) for each of the phase matched configurations in Fig. \ref{fig:Modes}(a), assuming no propagation loss.}
	\label{fig:CEvsPhi}
\end{figure}

Using the above formalism, we calculate the theoretical normalized conversion efficiency in units of $\%$ W$^{-1}$cm$^{-2}$ for each of the phase matched configurations in Fig. \ref{fig:Modes}(a), which we label as H and V according to the input pump polarization. Note that normalized conversion efficiency is defined $\frac{P_{2\omega}}{(P_{\omega} \cdot L)^2}$ for SHG and is simply given by $|\kappa|^2$ in the case of exact phase matching. We set $\theta = 15^\circ$, as in our thin film GaP sample, and vary the $\varphi$ angle. See Fig. \ref{fig:Experimental_crystalaxis}(c) for definitions of $\theta,\varphi$. The results are shown in Fig. \ref{fig:CEvsPhi}, with the conversion efficiency being strongly dependent on $\varphi$. In particular, a maximum conversion efficiency of 6780 $\%$ W$^{-1}$cm$^{-2}$ is predicted for V pump polarization and 356 $\%$ W$^{-1}$cm$^{-2}$ for H pump polarization, at $\varphi = 90^\circ$. These values are comparable to or even better than the state of art for LiNbO$_3$ platform \cite{Wang_2017OE,Chang_2016,Boes_2019}.

\section{Fabrication of the GaP nano-waveguides}
\label{sec:fab}

\begin{figure}
	\centering
	\includegraphics[trim={170 50 180 50},clip,width=0.8\linewidth]{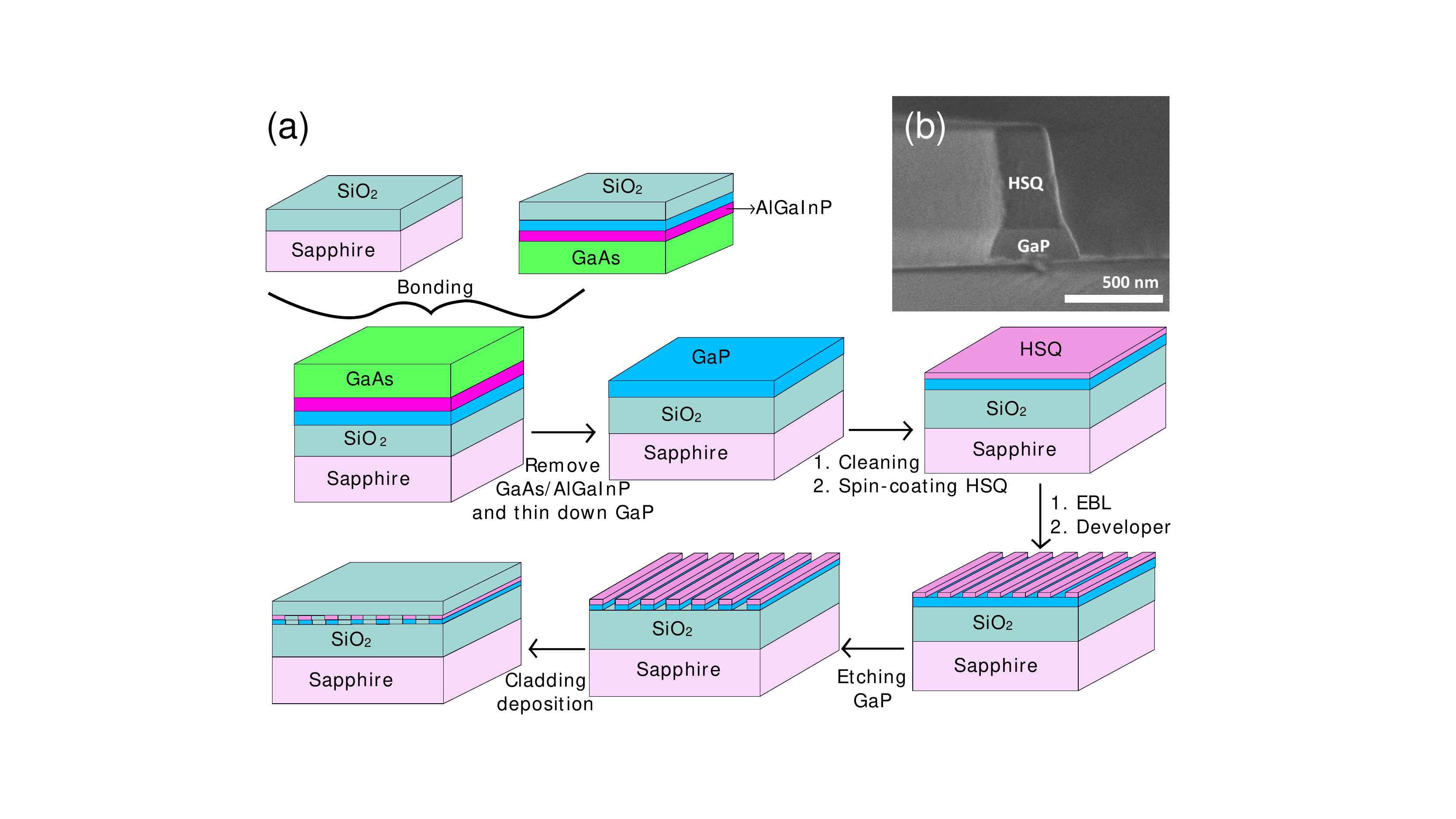}    
	\caption{(a) Fabrication flow, refer to text for details. (b) SEM image of fabricated waveguide with height of 215 nm and side wall angle of 70$^{\circ}$. The deviation from the nominal dimensions is due to fabrication uncertainty.}
	\label{fig:Fabrication_SEM}
\end{figure}

We start by depositing an AlGaInP buffer layer on a GaAs substrate by metal-organic chemical vapor deposition (MOCVD). This is to reduce the lattice mismatch between GaAs and GaP. Then a crystalline GaP active layer of thickness $\sim$400 nm was grown on top. This structure is directly bonded to a 150 $\mu$m sapphire substrate after depositing $\sim$2 $\mu$m SiO$_2$ layers on the top of both bonding surfaces. The GaAs substrate is then removed by wet etching, as shown in Fig. \ref{fig:Fabrication_SEM}. Finally, the wafer is cut into square samples. Each has a GaP layer of about 400 nm on top of few $\mu$m of SiO$_2$, which sits on top of the sapphire substrate. 

The fabrication of the GaP nano-waveguides begins with a standard wafer cleaning procedure (using acetone, iso-propyl alcohol and deionized water in that sequence under sonication). GaP layer was thinned down to the designed thickness of 220 nm using inductively-coupled plasma reactive ion etching (ICP-RIE) with N$_2$ and Cl$_2$ gas. The sample was followed by O$_2$ and hexamethyl disilizane (HMDS) priming in order to increase the adhesion between GaP and subsequent spin-coated electron-beam lithography (EBL) resist of hydrogen silsesquioxane (HSQ). After spin-coating of HSQ layer with a thickness of $\sim$540 nm, EBL and development in 25$\%$ tetra-methyl ammonium hydroxide (TMAH) defines the nano-waveguide regions in HSQ. ICP-RIE is then used to transfer the HSQ patterns to GaP. Finally, $\sim$3.2 um SiO$_2$ cladding layer is deposited on top of the waveguides by ICP-CVD. Devices are sent to a vendor to be diced by laser cutting. 

\subsection{Fabricated waveguide dimensions}
The fabricated waveguides have a height of 220 nm, with the top width varying from 280 nm to 420 nm in steps of 10 nm and a side wall angle of $70^\circ$. Due to the non-uniformity of the thickness of the GaP thin film ($\pm 5$ nm) across the wafer and the inevitable deviations in the fabricated waveguide width and side wall angle, there will be uncertainty in the final fabricated waveguide dimensions. The waveguide dimensions above were chosen to ensure that, within the expected uncertainty range of the final dimensions, we would observe SHG with our tunable O-band pump laser. The waveguides have a length of 1.5 mm excluding tapered edge couplers. The tapered edge couplers help with mode coupling from the lensed fiber to the waveguide. They have a length of 100 $\mu$m at each facet, with the starting top width of 220 nm. The cross-sectional scanning electron microscope (SEM) image of the waveguide is given in Fig. \ref{fig:Fabrication_SEM}(b). 

Our waveguides are fabricated to have a propagation direction oriented perpendicular to the edge of the square samples cut from the wafer. As shown in the next section, we observe SHG for both H and V polarized pump. From Fig. \ref{fig:CEvsPhi}, the conversion efficiency for H pump at $\varphi = 0^\circ$ is nearly zero. This indicates that our waveguides are oriented with $\varphi = 90^\circ$.


\section{Measurement results}

\subsection{Experimental setup to study SHG}

\begin{figure}
	\centering
	\includegraphics[width=0.75\linewidth]{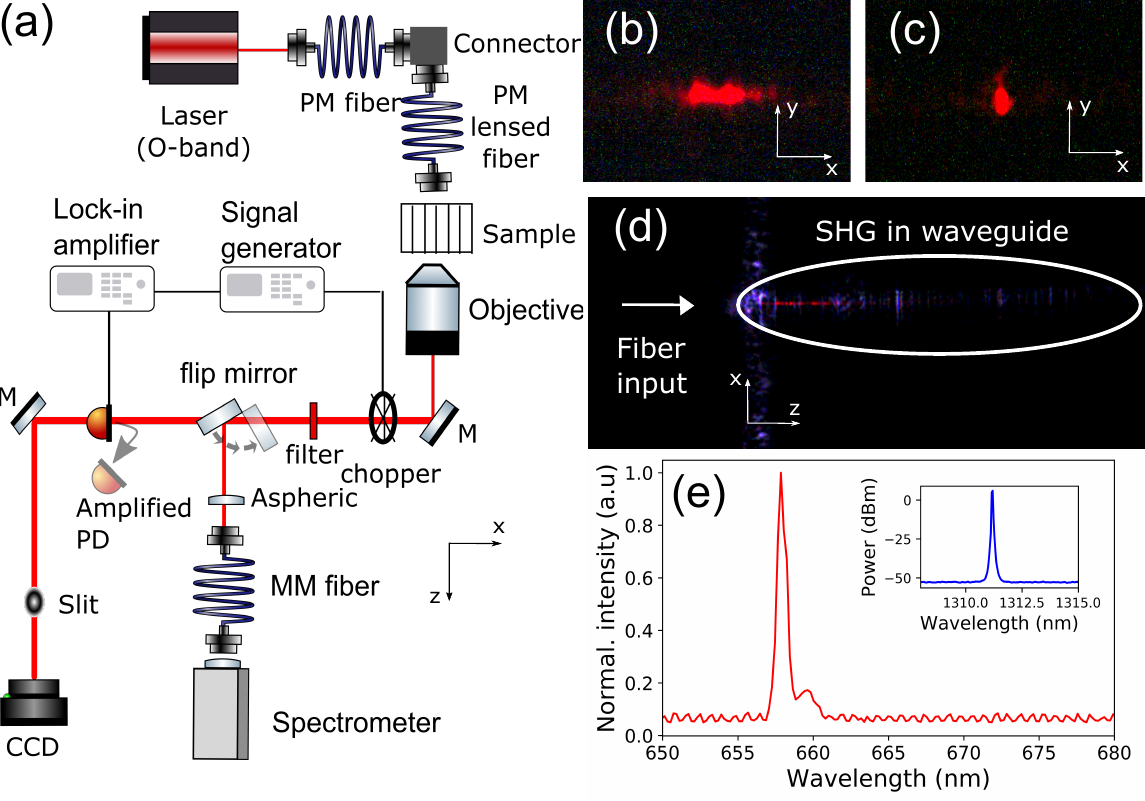}
	\caption{(a) Experimental setup showing tapered fiber input, chip mount, output objective to collect the second harmonic light, CCD camera used to analyze the second harmonic. CCD--camera, MM--multi mode, M--mirror, PM--polarization maintaining, PD--photodetector; (b) SHG output from waveguide on the CCD camera for H polarized input pump; (c) SHG output from the waveguide on the CCD camera for V polarized input pump; (d) Top view of the waveguide when pump light is coupled in, with the waveguide marked using an ellipse where the SHG generation can be observed; (e) Spectrum of SHG with the  pump spectra given in the inset.}
	\label{fig:setup}
\end{figure}

Figure \ref{fig:setup}(a) shows the schematic of the experimental setup\cite{Optics}. The pump light from a narrow linewidth continuous-wave tunable laser (Yenista, 1260 nm - 1360 nm, linewidth $\sim$400 kHz) is coupled into the waveguide using a tapered polarization-maintaining lensed fiber (OZ optics) designed for wavelengths around 1310 nm. An objective lens with a numerical aperture (NA) of 0.7 (Mitutoyo) is used to collect the higher order second  harmonic light output from the waveguide. The output from the objective is imaged using a charged coupled device (CCD) camera (uEye), which also gives the intensity readings in counts. The CCD camera is preceded by a filter centered at 650 nm and having a bandwidth of 150 nm (Semrock). Another 10x objective lens and CCD camera are mounted above the structure to observe the interface with the fiber. A flip mirror is used to direct the SHG light into the spectrometer (Ocean Optics). The lock-in amplifier (Signal Recovery) is phase-locked to the chopper rotating at a frequency of 500 Hz, which is used for modulating the SHG signal, and to the amplified photo-detector (PD, Thorlabs) to detect the SHG signal. Since the chopper reduces the power by half, it is introduced at the output of the waveguide for SHG instead of the input pump because of the quadratic dependence between the pump power and the SHG power. SHG is studied in the waveguide for H and V pump polarizations. The polarization of the pump is set by rotating the tapered lensed fiber and testing the output from the fiber through a polarizer without the waveguides in the path.

\subsection{Experimental observations and analysis}
First, we observe the mode structure of the SHG by using the CCD camera at the output of the chip. Figure \ref{fig:setup}(b) shows the SHG mode for the H polarized pump at 1330.7 nm with the top width of 350 nm. Figure \ref{fig:setup}(c) shows the SHG mode for the V polarised pump at 1359.3 nm with the top width of 310 nm. It should be noted that the measured mode profiles presented in Fig. \ref{fig:setup}(b, c) are obtained after the tapered edge coupler, which affects the output mode profile of the SHG. Figure \ref{fig:setup}(d) shows the image from the top CCD camera when pump light is coupled into the waveguide and there is SHG scattered by the waveguide. Figure \ref{fig:setup}(e) shows the spectrum of the SHG obtained from the spectrometer and the inset shows the optical spectrum of the pump light.   

\begin{figure}
	\centering
	\includegraphics[trim={40 0 30 0},clip,width=1\linewidth]{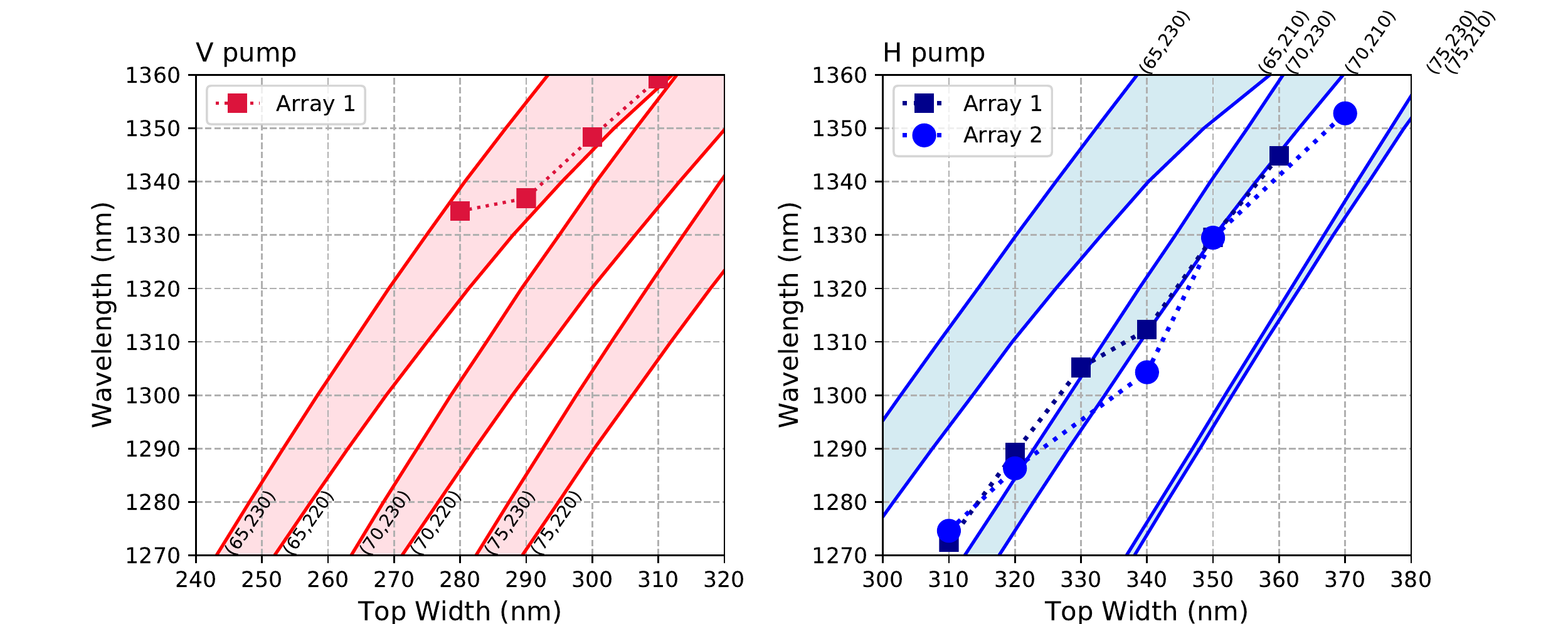}
	\caption{Experimental (markers) and simulated (lines and shaded regions) phase matched second harmonic wavelengths in waveguides for H and V polarized input pump light. Labels indicate the (side wall angle, height) used in simulation. Each shaded region represents a fixed side wall angle with continuously varying height.}
	\label{fig:Phasematching}
\end{figure}

Next we study the dependence of the phase matching SHG wavelengths on the waveguide top width. We fabricated arrays of waveguides, with the top width increasing in steps of 10 nm. We tune the pump wavelength from 1260 nm to 1360 nm in steps of 0.1 nm and take the phase matching wavelength as the value where SHG power is maximum. Figure \ref{fig:Phasematching} shows the dependence of the phase matched pump wavelength on the top width of the waveguide for V (red markers) and H (blue markers) polarized input pump light. The phase matching wavelengths varies between nominally identical structures due to uncertainty in the dimensions of the fabricated structures. 


To verify that we are observing SHG due to the interacting modes depicted in Fig. \ref{fig:Modes}, we compare the experimentally measured phase matching wavelengths with simulation results. Fabrication uncertainty was accounted for by simulating waveguides with side wall angle of $65^\circ$ to $75^\circ$ and height from 210 nm to 230 nm. Each shaded region in Fig. \ref{fig:Phasematching} represents a fixed side wall angle with continuously varying height, as indicated by the labels. For clarity, we do not plot heights less than 220 nm for V polarization as the shaded regions begin to overlap. We observe a trend towards smaller side wall angle for narrower top widths. Overall, the experimental results agree well with the simulations, affirming that we have identified the correct interacting modes. 


\begin{figure}
	\centering
	\includegraphics[width=1\linewidth]{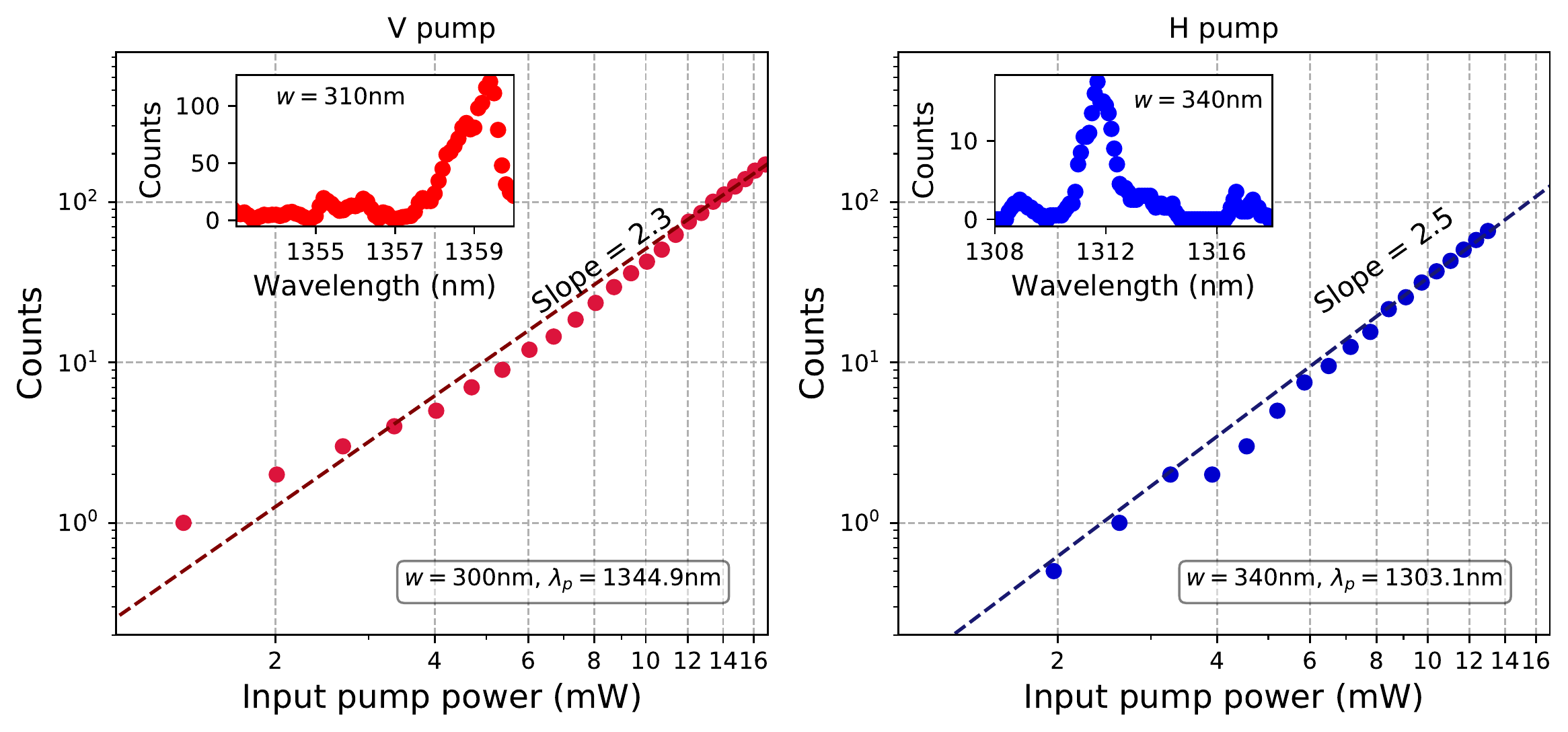}
	\caption{Dependence of output second harmonic counts (dots) on the input pump power into the nano-waveguide. Log-linear fit to the experimental data is given by dotted lines. (Insets) Counts versus wavelength, showing the phase matching bandwidth. }
	\label{fig:ConversionEfficiency_curvefit}
\end{figure}

Figure \ref{fig:ConversionEfficiency_curvefit} shows the dependence of the SHG as a function of input pump power. The pump polarization, wavelengths and waveguide widths are as labeled in the plot. As we reduced the pump power, we reached the detection limit of the amplified PD. We thus used the counts on the CCD as a measurement of the SHG power. The circles (dotted lines) represent the experimental data (fit). The SHG intensity on the CCD varies as a function of the input pump power to the waveguide with a slope of approximately two in the log-scale, as expected from the theory. The deviation from ideal quadratic scaling is likely due to a slight nonlinearity in CCD response. The insets of Fig. \ref{fig:ConversionEfficiency_curvefit} show the counts versus the pump wavelength, giving a phase matching bandwidth of $\sim$1.2 nm (FWHM) for both H and V pump polarizations. The simulated phase matching bandwidth is  $\sim$0.2 nm and we attribute the broadening to non-uniformities along the waveguide. 


Since coupling of higher order modes is experimentally challenging, we were not able to measure the total transmission for the SHG modes. We thus present the external SHG conversion efficiency. A high power CW pump laser (Hubner), tuned to a wavelength of 1311.8 nm, was used. At the end of the input lensed fiber, the pump power was 160 mW. The H polarized pump was then input into a waveguide of top width 340 nm. The SHG power at the output of the waveguide, measured with amplified PD and lock-in amplifier, was 773 pW. Thus, the normalized external conversion efficiency is $\frac{P_{2\omega}}{(P_\omega \cdot L)^2} = 1.34 \times 10^{-4}$ $\%$ W$^{-1}$cm$^{-2}$. As the high power pump laser has a wavelength tuning limit of about 1330 nm, we did not measure the SHG efficiency with V polarization. However, by comparing the CCD counts in Fig. \ref{fig:ConversionEfficiency_curvefit}, we deduce that SHG with V polarized pump does indeed have higher conversion efficiency. 

We calculate the theoretical internal normalized conversion efficiency for the H pump configuration to be 356 $\%$ W$^{-1}$cm$^{-2}$. The fiber-to-waveguide coupling loss and the high propagation loss contribute to the mismatch between the experimental and theoretical conversion efficiencies. The total transmission for pump wavelengths is approximately $-10$ dB for H polarization. After factoring the coupling losses of about $-7$ dB, we estimate the pump propagation losses to be 2 dB/mm. We attribute the high propagation losses to the large refractive index contrast between GaP and the SiO$_2$ cladding, as well as roughness of the waveguide interfaces \cite{Tien_1971}. Moreover, we expect the propagation loss at the second harmonic wavelength to be higher since roughness scattering loss $\propto 1/\lambda^2$. To model the effects of losses, Eq. \ref{eq:pump-pm} and \ref{eq:SHG-pm} can be augmented by adding loss terms. Assuming undepleted pump and exact phase matching, we can derive an effective length of interaction, 
\begin{equation}
L_{\textrm{eff}} = \frac{e^{-\alpha_\omega L}-e^{-\alpha_{2\omega} L/2}} {\frac{\alpha_{2\omega}}{2}-\alpha_\omega}.
\end{equation} 
The internal normalized conversion efficiency is then reduced by a factor $(L_{\textrm{eff}}/L)^2$. Another cause for the discrepancy between experiment and theory is the non-uniformity of the waveguide dimensions along its length due to fabrication imperfections.  Variations in geometry causes shifts in the phase matching wavelength, as seen from Fig. \ref{fig:Phasematching}, and disrupts the coherent build-up of the second harmonic signal. 

Several recent demonstrations of SHG using resonant GaP structures have shown good conversion efficiencies \cite{Rivoire_2009,Rivoire_2011,Lake_2016,Logan_2018}. However, for our results in waveguided structures, there is a discrepancy between calculated and measured SHG, due to fabrication limitations. An ongoing effort is underway to improve the fabrication uniformity (surface roughness and dimension control) and optimize the device design, in order to achieve higher experimental conversion efficiencies. 

\section{Conclusion}

Frequency conversion between the visible and telecom wavelength regions is needed to interface single photon emitters with optical fibers, a critical step for long distance quantum communications technology \cite{Greve_2012,Zaske_2012}. GaP has many advantageous material properties, such as high nonlinearity and cut-off wavelength of 550 nm, which are very well suited for such nonlinear and quantum optics applications. 

We have reviewed the theory of SHG in GaP nano-waveguides, taking into account the configuration of the nonlinear susceptibility tensor. Prior to processing the GaP thin film, we measured and confirmed the crystal axis orientation, which is critical for the modeling of SHG. We designed, fabricated and tested gallium phosphide nano-waveguides for SHG. The fabricated devices of different widths generated second harmonic light around 655 nm for low power continuous-wave input pump wavelengths. We observed SHG interaction between different modes depending on the pump polarization, pump wavelength and waveguide width. The dependence of experimentally obtained phase matched wavelengths on the waveguide top width closely matches with the simulations. 

Efforts are in progress to increase experimental conversion efficiency through improved fabrication and optimized waveguide designs. The results presented in this work will be useful for the further development of nonlinear and quantum photonic platforms based on GaP. 

\section*{Funding}

Agency for Science, Technology and Research (A*STAR), Science and Engineering Research Council (SERC), Quantum Technologies for Engineering (QTE), (A1685b0005).

\section*{Acknowledgments}
We thank Victor Leong for his help in designing and building the experimental systems.

\section*{Disclosures}
The authors declare no conflicts of interest.


\bibliography{SHG_References}

\end{document}